\pdfoutput=1
\documentclass[aps,prl,preprintnumbers,amsmath,amssymb,nofootinbib,superscriptaddress,floatfix,showkeys,twocolumn]{revtex4-1}

\usepackage{graphicx}
    \graphicspath{{Figs/}}
\usepackage{xcolor}

\usepackage{hyperref}
\usepackage{slashed}
\usepackage{soul}
\usepackage[normalem]{ulem}
\usepackage[caption=false]{subfig}
\usepackage{booktabs}
\usepackage{diagbox}
\usepackage{makecell}
\usepackage{multirow}
\usepackage{enumitem}

\usepackage{physics}
\usepackage[compat=1.1.0]{tikz-feynhand}

\makeatletter
\pretocmd{\@hang@from}{\noindent}{}{}
\pretocmd{\@hang@froms}{\noindent}{}{}
\pretocmd{\@hangfrom@appendix}{\noindent}{}{}
\pretocmd{\@hangfroms@appendix}{\noindent}{}{}
\makeatother

\newcommand{\lsim}{\mathrel{\mathop{\kern 0pt \rlap
  {\raise.2ex\hbox{$<$}}}
  \lower.9ex\hbox{\kern-.190em $\sim$}}}
\newcommand{\gsim}{\mathrel{\mathop{\kern 0pt \rlap
  {\raise.2ex\hbox{$>$}}}
  \lower.9ex\hbox{\kern-.190em $\sim$}}}
\newcommand{\sv}{\ensuremath{\langle\sigma v_{\text{rel}}\rangle}}

\newcommand{\kev}{\ensuremath{\,\mathrm{keV}}}

\newcommand{\gev}{\ensuremath{\,\mathrm{GeV}}}

\DeclareDocumentCommand{\lettersection}{m}{\textit{#1}---\ignorespaces}

\def  \bcen   {\begin{center}}
\def  \ecen   {\end{center}}
\def  \beq    {\begin{equation}}
\def  \eeq    {\end{equation}}
\def  \beqa   {\begin{eqnarray}}
\def  \eeqa   {\end{eqnarray}}

\def\bea{\begin{eqnarray}}
\def\eea{\end{eqnarray}}

\begin{document}

\title{
A Common $\sim$55 GeV Inelastic Dark Matter Origin of the LZ, Fermi-LAT, and AMS-02 Excesses and Paleo-Detector Prospects}

\author{Meiwen Yang}
\affiliation{Key Laboratory of Dark Matter and Space Astronomy, Purple Mountain Observatory, Chinese Academy of Sciences, Nanjing 210033, China}
\affiliation{School of Astronomy and Space Science, University of Science and Technology of China, Hefei, Anhui 230026, China}

\author{Quan-feng Wu}
\author{Kai-Kai Duan}
\author{Zi-Qing Xia}
\affiliation{Key Laboratory of Dark Matter and Space Astronomy, Purple Mountain Observatory, Chinese Academy of Sciences, Nanjing 210033, China}
\author{Yue-Lin Sming Tsai}
\author{Yi-Zhong Fan}
\affiliation{Key Laboratory of Dark Matter and Space Astronomy, Purple Mountain Observatory, Chinese Academy of Sciences, Nanjing 210033, China}
\affiliation{School of Astronomy and Space Science, University of Science and Technology of China, Hefei, Anhui 230026, China}

\date{\today}

\begin{abstract}
We investigate the 248~keV nuclear recoil excess reported by LUX-ZEPLIN (LZ) in a leptophobic inelastic dark matter framework, in which the mass splitting naturally suppresses low-energy recoils.
We find two viable branches, including the endothermic solution with a high dark matter mass $m_\chi \gtrsim400$~GeV
and the exothermic one with a mass of tens GeV.
Remarkably, the exothermic dark matter scenario with $m_\chi\sim55$~GeV can simultaneously account for the LZ excess,
the Galactic Center GeV gamma-ray excess, and the AMS-02 antiproton excess.
The favored annihilation cross-section is slightly below the one required for the observed relic abundance, resulting in an overabundant relic in the minimal setup.
An additional neutrino-portal mediator can provide the required depletion while leaving the gamma-ray and antiproton signals essentially unchanged.
We further show that paleo-detectors provide a powerful probe of the inelastic interpretation, requiring exposures of $\sim10$ and $\sim100$~mg Gyr for the endothermic and exothermic scenarios, respectively.
\end{abstract}

\maketitle

\lettersection{Introduction}
The LUX-ZEPLIN (LZ) collaboration has recently reported a search for dark matter interactions in an extended nuclear recoil energy window up to $E_r \simeq 270\kev$, motivated by effective field theory and inelastic dark matter models that predict non-negligible recoil spectra at higher energies \cite{LZ:2026axp}.
In a $2.84$ tonne-year exposure, the collaboration observes a single candidate event with a reconstructed nuclear recoil energy of $248 \pm 23\ (\text{stat}) \pm 23\ (\text{sys})\kev$, lying in a region of the $\{S1c, S2c\}$ plane where the background expectation is low. A profile likelihood ratio test yields a global significance of $2.6\sigma$ after accounting for look-elsewhere effects, with a maximum local significance of $3.4\sigma$ across the tested models.

This observation departs from the elastic spin-independent dark matter-nuclei scattering scenario,
in which the recoil spectrum is concentrated at low energies and rapidly decreases toward the high-energy tail.
A high-energy recoil without a corresponding low-energy excess can arise from non-standard dark-matter interactions.
In particular, endothermic or exothermic inelastic scattering modifies the recoil kinematics through a mass splitting $\delta$,
\begin{equation}
v_{\min}(E_r)=\frac{1}{\sqrt{2m_NE_r}}
\left(\frac{m_NE_r}{\mu_N}\pm \delta\right),
\end{equation}
where $+\delta$ (endothermic scattering) raises the velocity threshold and can suppress low-energy recoils,
whereas $-\delta$ (exothermic scattering) lowers $v_{\min}$ and can enhance the relative contribution of high-energy recoils.
Alternatively, velocity- or momentum-dependent interactions can reshape the spectrum through the scattering matrix element,
enhancing the relative rate at large recoil energies.
For xenon, these mechanisms can produce an appreciable rate at $E_r \simeq 248$~keV without triggering a low-energy excess.
This behavior holds even for light dark matter in exothermic scattering scenario,
whereas endothermic, velocity-, or momentum-dependent interactions generally require heavier dark matter, typically $m_\chi \gtrsim 100~\text{GeV}$.

The anomaly has prompted inelastic dark matter interpretations, yet the predictive thermal Higgsino benchmark faces severe tensions from solar-neutrino and high-energy sideband constraints \cite{2609.01583,Fan:2026kxx,2609.02775,2609.04175, Bisal:2026khf, Bose:2026ndd}.
Alternative UV models and scattering mechanisms highlight the non‑uniqueness of the explanation
\cite{2609.01475,2609.02608,2609.01892,2609.04163,2609.02807,2609.02505,2609.04144,2609.01592,2609.04186,2609.04185, deLima:2026shq, Dent:2026bji},
while gamma‑ray claims rely on distinct astrophysical assumptions and do not provide independent confirmation \cite{2609.01590,2609.02868,2609.02994}.
A convincing interpretation must instead predict a consistent multi‑messenger signal.
Future target‑dependence and seasonal modulation offer discriminative tests, unlike treating one June event as an annual cycle \cite{2609.04181}.
However, it remains unclear whether the 248 keV excess can be embedded in a single predictive framework that simultaneously satisfies direct-detection,
indirect-detection, and cosmological constraints.

Here we show that the 248 keV LZ excess can arise from a predictive leptophobic inelastic dark matter framework with consistent multimessenger signatures.
In particular, the exothermic scenario around $m_\chi\sim55$ GeV provides a common origin for the LZ excess,
the Galactic Center gamma-ray excess (GCE)~\cite{Hooper:2010mq,Daylan:2014rsa,Zhou:2014lva,Calore:2014nla,Zhong:2019ycb},
and the AMS-02 antiproton excess~\cite{Cui:2016ppb,Cuoco:2016eej},
while the endothermic scenario at $m_\chi\gtrsim400$ GeV does not reproduce the Galactic Center spectrum.
The resulting inelastic kinematics and associated indirect signals yield complementary tests in cosmology, neutrino observations, and paleo-detectors.

\begin{figure*}[!ht]
\includegraphics[width=0.49\textwidth]{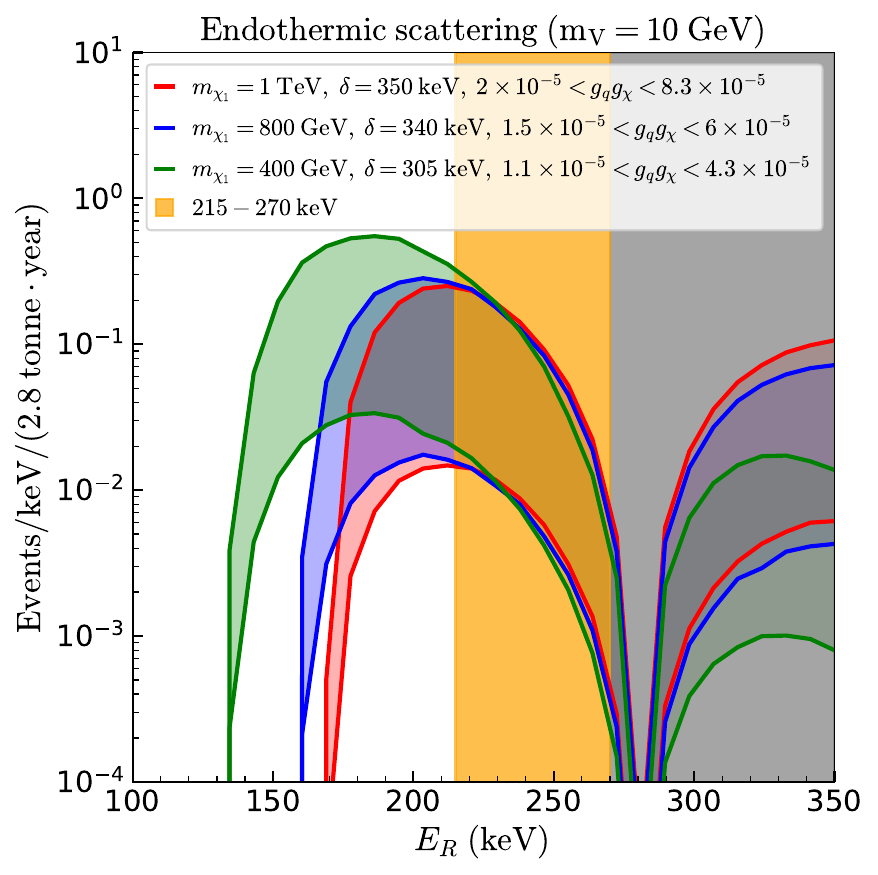}
\includegraphics[width=0.49\textwidth]{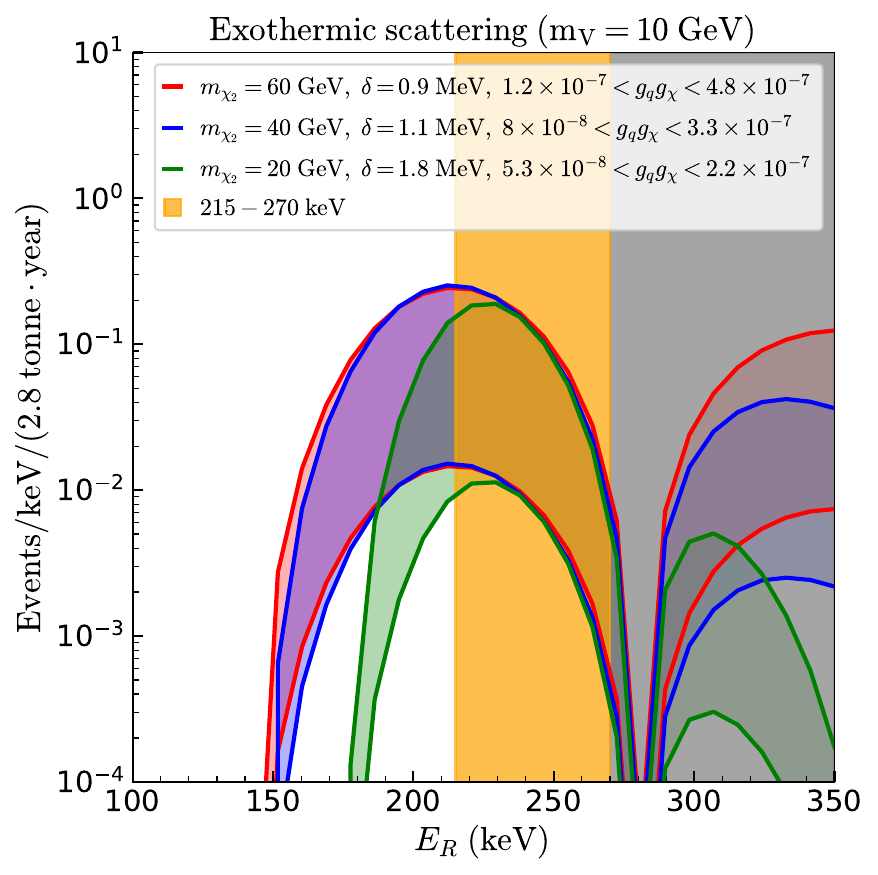}
\caption{Benchmark spectra reproducing the LZ event for endothermic (left) and exothermic (right) scattering with $m_V = 10~\mathrm{GeV}$.
Each band spans the $90\%$ C.L. range of $g_q g_\chi$ given in the legend; the exothermic masses are those of $\chi_2$.
The orange band marks the $215\text{--}270\kev$ window, and the gray region marks recoil energies above $269.9\kev$, at which the LZ efficiency falls to $50\%$~\cite{LZ:2026axp}.
}
\label{fig:signal}
\end{figure*}

\lettersection{Model and parameter space}
We consider a minimal leptophobic realization of inelastic dark matter \cite{Tucker-Smith:2001myb}: a Dirac fermion split by small Majorana masses into two Majorana states, $\chi_1$ and $\chi_2$, with $\delta \equiv m_{\chi_2} - m_{\chi_1} \ll m_{\chi_1}$.
It interacts with quarks through a massive vector boson $V_\mu$ with mass $m_V$,
\begin{equation}
    \mathcal{L}_{\mathrm{int}} = g_q V_\mu \sum_q \bar{q} \gamma^\mu q + i g_\chi V_\mu \bar{\chi}_2 \gamma^\mu \chi_1,
    \label{eq:interaction}
\end{equation}
where the sum runs over all six quark flavors.
The quark current is three times the baryon-number current, so the nucleon coupling is exactly $3 g_q$.
Such a coupling arises, for example, when $V_\mu$ gauges a spontaneously broken baryon number under which the dark fermion is charged \cite{Lee:1955vk, Carone:1994aa, Tulin:2014tya, Gemmell:2026yaw}.
The same field content underlies the exothermic interpretation of the LZ event in Ref.~\cite{Gemmell:2026yaw}, whereas Ref.~\cite{Du:2026lpa} gauges baryon number with chiral dark charges for an endothermic one.
The diagonal vector bilinears $\bar{\chi}_i \gamma^\mu \chi_i$ of Majorana fermions vanish identically \cite{Kayser:1982br, Nieves:1981zt}, and for equal Majorana masses of the two Weyl components no diagonal axial-vector coupling arises, leading to $V_\mu$ coupling only to the transition $\chi_1 \leftrightarrow \chi_2$.
Scattering on nuclei is therefore purely inelastic, spin independent, and isospin conserving.
For $m_V < m_{\chi_1}$, dark matter annihilates dominantly through $\chi_i \chi_i \to V V$, with
\begin{equation}
    \sigma v = \frac{g_\chi^4}{16 \pi m_{\chi_1}^2} \frac{(1 - r)^{3/2}}{(1 - r/2)^2}, \qquad r \equiv \frac{m_V^2}{m_{\chi_1}^2},
    \label{eq:sigmav}
\end{equation}
while the purely off-diagonal coupling forbids $\chi_1 \chi_2 \to V V$.
For $m_V = 10~\mathrm{GeV}$, $V$ decays almost exclusively into $q \bar{q}$.
These properties and the estimates below are derived in the Supplemental Material.

Beyond Eq.~\eqref{eq:interaction}, the only renormalizable operators that mix $V_\mu$ with the neutral gauge bosons are kinetic mixing with hypercharge \cite{Holdom:1985ag}, which below the electroweak scale appears as a kinetic mixing $\varepsilon$ with the photon, and mass mixing with the $Z$ boson, $m_{VZ}$ \cite{Babu:1997st, Davoudiasl:2012ag}.
Quark loops generate $\varepsilon$, but it stops running below $m_c$, where the charges of $u$, $d$, and $s$ sum to zero.
We take the low-energy mixing $\varepsilon \equiv \varepsilon(\mu \lesssim m_c)$ to be small, a condition that is stable under renormalization below $m_c$.
The running above $m_c$ then gives $\abs{\varepsilon(m_V)}$ of a few percent of $g_q$ and $\mathrm{BR}(V \to \ell^+ \ell^-) \lesssim 10^{-4}$, which is harmless.
The conserved quark current does not generate the mass mixing, and if $V_\mu$ gauges baryon number, $m_{VZ}$ can arise only from states that carry both charges and obtain mass from electroweak symmetry breaking, such as the fermions that cancel the anomalies of a gauged baryon number \cite{FileviezPerez:2010gw, Dobrescu:2014fca, Dobrescu:2021vak}.
For $m_V = 10~\mathrm{GeV}$ and the couplings below, the anomaly-enhanced processes of Ref.~\cite{Dror:2017ehi} are negligible.

The two mixings control the fate of $\chi_2$.
For $\delta < 2 m_e$, $\chi_2$ decays only to $\chi_1 + 3 \gamma$ or $\chi_1 \nu \bar{\nu}$, and it survives to the present if $m_{VZ}$ lies below a few MeV, as we assume following Ref.~\cite{Gemmell:2026yaw}.
For $\delta > 2 m_e$, the low-energy $\varepsilon$ opens $\chi_2 \to \chi_1 e^+ e^-$, whose width, once the LZ rate is fixed, depends on $\varepsilon / g_q$ but not on $m_V$, as long as $m_V$ exceeds the momentum transfer of about $250~\mathrm{MeV}$ in the LZ event.
The positrons from these decays must not exceed the Galactic 511 keV emission \cite{Siegert:2015knp}, which requires $\varepsilon / g_q \lesssim 10^{-3}$ and $3 \times 10^{-5}$ together with $m_{VZ} \lesssim 2$ and $0.5~\mathrm{MeV}$ for the benchmarks with $\delta = 1.1$ and $1.8~\mathrm{MeV}$.
These bounds on $\varepsilon$ lie at least an order of magnitude below the value $(0.02\text{--}0.06)\, g_q$ that loops of anomaly-cancelling fermions with masses of $0.1$--$1~\mathrm{TeV}$ would generate.
The opposite regime, $m_{VZ} \gtrsim 13\text{--}20~\mathrm{MeV}$, empties the heavy state and instead admits an endothermic interpretation.
Since $\delta$ lies far below the freeze-out temperature, both states freeze out with equal abundances,
giving $\langle \sigma_{\mathrm{eff}} v \rangle = \sigma v / 2$ and fixing $g_\chi$ through the relic density.
Perturbativity, $g_\chi \lesssim 1$, then requires $m_{\chi_1} \lesssim 2.3~\mathrm{TeV}$ for $m_V = 10~\mathrm{GeV}$.

After freeze-out, dark conversion $\chi_2 \chi_2 \to \chi_1 \chi_1$ through $t$- and $u$-channel $V$ exchange lowers the heavy-state fraction to $f_2 = 0.42 \text{--} 0.49$ for the exothermic benchmarks \cite{Duan:2026zqj}.
Its rate scales as $g_\chi^4 / m_V^4$, so $f_2 \simeq 1/2$ for the $m_V = 30~\mathrm{GeV}$ of Ref.~\cite{Gemmell:2026yaw}.

The parameters thus separate into two groups.
First, the LZ spectrum selects $(m_{\chi_1}, \delta)$, and the LZ rate fixes $(g_q g_\chi)^2$ times the abundance of the incoming state,
namely $f_2$ for exothermic and $1 - f_2 \approx 1$ for endothermic scattering.
Second, in the minimal setup the thermal history fixes $g_\chi$ and $f_2$ for given $(m_{\chi_1}, \delta, m_V)$, so $g_q$ follows from the LZ rate and affects only the experimental constraints discussed below.

Fixing $m_V=10~\mathrm{GeV}$, we select six benchmark points from the $90\%$ C.L. signal regions of Ref.~\cite{LZ:2026axp}, guided by the LZ recoil spectra.
The corresponding spectra for the endothermic (left) and exothermic (right) benchmarks are shown in Fig.~\ref{fig:signal},
and their parameter details are given in the Supplemental Material.
The required $g_q$, between $5.5 \times 10^{-7}$ and $1.3 \times 10^{-4}$, lies orders of magnitude below the limits from dijet, electroweak, and quarkonium measurements \cite{Dobrescu:2021vak, CMS:2019xai}.
It still makes $V$ decay within $10^{-12}~\mathrm{s}$, much faster than the Hubble expansion at freeze-out, so the dark sector stays in equilibrium with the SM, as assumed above.

\lettersection{Astrophysical signatures of dark matter}
With the scattering parameters constrained by the LZ signal regions,
we use the GCE and AMS-02 antiproton data as complementary probes of the remaining parameter space.
Using the Lagrangian in Eq.~\eqref{eq:interaction}, we calculate the gamma-ray and antiproton fluxes from $\chi_1\chi_1\to 2V$ and $\chi_2\chi_2\to 2V$ annihilation, together with the associated neutrino flux.
Nonrelativistic annihilation is kinematically forbidden for $m_V>m_{\chi_1}$.
For the mass range selected by the LZ event, the endothermic branch produces annihilation spectra that cannot account for either excess.
We therefore focus on the exothermic branch in the following indirect-detection analysis, while presenting the corresponding endothermic results in the Supplemental Material.

For the leptophobic mediator with $m_V=10\gev$, dark matter annihilates via $\chi_{1,2}\chi_{1,2}\to VV$,
followed by the on-shell decay $V\to q\bar q$, resulting in a four-quark final state.
Since the mediator decays on shell, the annihilation cross section $\sv$ is independent of the quark coupling $g_q$.
The flux of a stable Standard Model particle $i=\gamma,\nu$ is then
\begin{equation}
\frac{d\Phi^{\rm ann}_i}{dE_i} \qty(\chi\chi\to V V)= \frac{\langle \sigma v_{\rm rel.} \rangle}{8\pi m_\chi^2} \sum_q \frac{dN_i^q}{dE} \, \mathcal{J},
\end{equation}
where $dN_i^q/dE_i$ denotes the photon or neutrino spectrum from $\chi_{1,2}\chi_{1,2}\to VV\to 4q$ using \texttt{PPPC4} \cite{Cirelli:2010xx} as performed in Ref.~\cite{Abdughani:2021oit}.
The astrophysical factor $\mathcal{J}$ encodes the halo distribution and propagation effects specific to each species,
$\mathcal{J}(\Delta\Omega) \equiv \int_{\Delta\Omega} d\Omega \int ds\, \rho^2(r)$, with a gNFW profile (slope index $1.28$ and local density $0.3$~GeV/cm$^3$).
With the LZ-favored $\delta$, $\chi_2$ remains sufficiently long-lived to account for about 50\% of the relic density after freeze-out,
as determined using the Boltzmann solver \cite{Duan:2026zqj}.

For the AMS-02 antiproton excess \cite{Cui:2016ppb,Cuoco:2016eej}, we compute the antiproton flux at Earth by propagating the dark matter-induced flux from the Galactic halo using \texttt{GALPROP}~\cite{1998ApJ...509..212S, 2007ARNPS..57..285S}.
Following the framework of Ref.~\cite{Duan:2025ead}, we apply its uncertainties in cosmic-ray propagation, background antiproton production,
and solar modulation, to test the compatibility of leptophobic dark matter annihilating dominantly into quarks with the AMS-02 antiproton \cite{2025PhRvL.134e1002A} excess around ten~GV.

\begin{figure*}[!ht]
\includegraphics[width=0.49\textwidth]{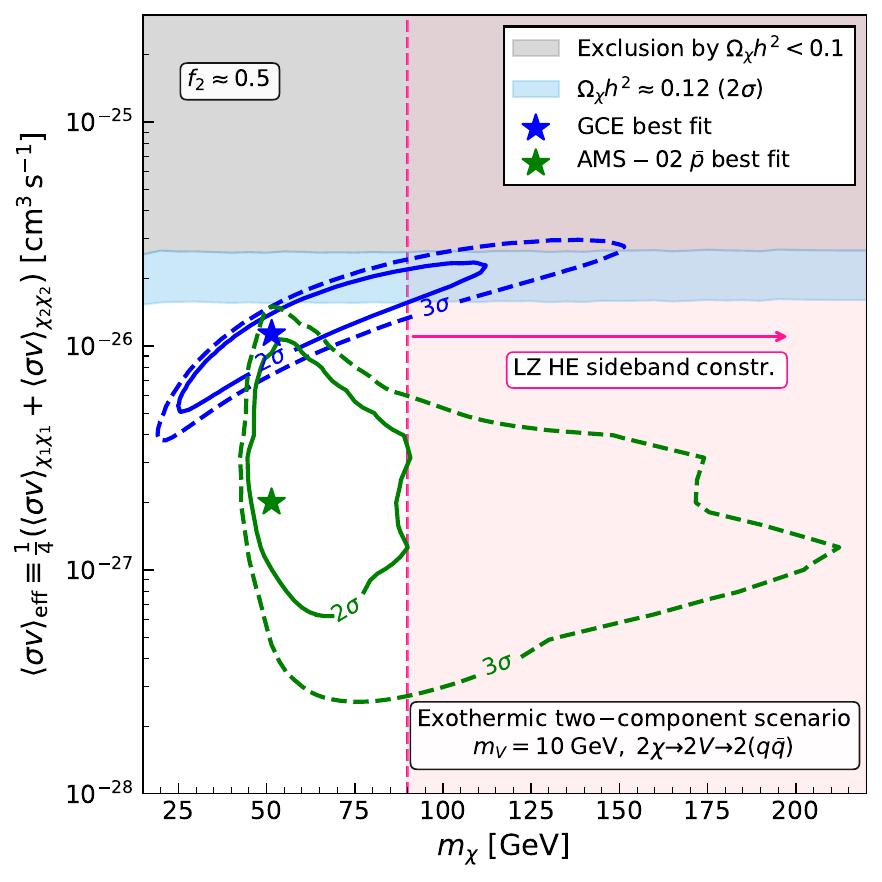}
\includegraphics[width=0.49\textwidth]{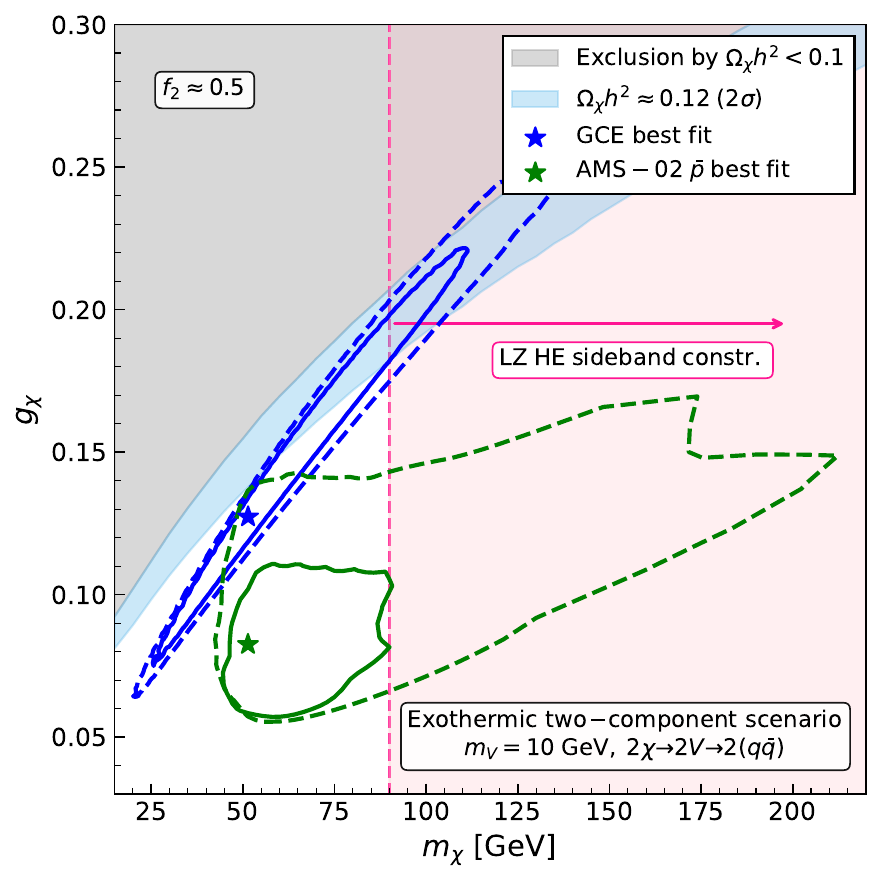}
\caption{Viable parameter space for the exothermic two-component dark matter scenario with equal present-day abundances of the two species.
The annihilation processes are $\chi_1\chi_1\to VV$ and $\chi_2\chi_2\to VV$, with each on-shell $V$ subsequently decaying into a quark pair,
resulting in a four-quark final state.
The gray region is excluded by relic-density constraints, corresponding to upper limits on $\langle\sigma v\rangle$ (left) and $g_\chi$ (right)
in the minimal-interaction setup.
The light blue shaded region yields the observed relic density in the minimal interaction setup.
Blue and green contours show the $2\sigma$ (solid) and $3\sigma$ (dashed) regions favored by the GCE and antiproton excess, respectively.
For $m_\chi > 90~\mathrm{GeV}$ (the pink region), the constraint from the LZ high-energy sideband relies on an extrapolation of the nuclear-recoil response beyond the official LZ search region~\cite{Dent:2026bji}.
}
\label{fig:mx_gx}
\end{figure*}

Figure \ref{fig:mx_gx} shows the viable parameter space for an exothermic scenario with two-component dark matter of equal present-day abundances.
The gray region is excluded by relic-density constraints, $\Omega_\chi h^2<0.1$.
In our minimal-interaction setup, these bounds serve as upper limits on $\langle\sigma v\rangle$ (left) and $g_\chi$ (right).
Introducing additional interactions would allow smaller values while still reproducing the correct relic density.
The blue and green contours denote the regions favored by the GCE and the antiproton excess, respectively.
Solid and dashed lines correspond to $2\sigma$ and $3\sigma$.
For the GCE contours, we carry out a $\chi^2$ analysis using the SED data and
the $\chi^2$ function given in the appendix of Ref.~\cite{Zhu:2022tpr}.
For $m_\chi \gtrsim 90~\mathrm{GeV}$ (pink region), the parameter space can be excluded
when extrapolating the nuclear-recoil response beyond the official LZ search region,
$E_r\gtrsim 350~\mathrm{keV}$, under the high-energy sideband constraints~\cite{deLima:2026shq}.

Interestingly, the region that accommodates the LZ anomaly, GCE, and antiproton excesses lies around $m_\chi\sim 55~\mathrm{GeV}$,
with $\sv\approx10^{-26}$~cm$^3$s$^{-1}$ and $g_\chi\approx0.13$.
However, its relic abundance exceeds the Planck measurement,
indicating that additional interactions of comparable strength are required to reduce the dark matter abundance.

Although both endothermic and exothermic scattering can account for the LZ anomaly,
the endothermic scenario requires $m_\chi\gtrsim400~\mathrm{GeV}$ and cannot accommodate the GCE spectrum.
The corresponding results for the endothermic scenario are provided in the Supplemental Material for completeness.
In addition, we check neutrino limits from the Galactic Center~\cite{IceCube:2014stg} and gamma-ray constraints from dwarf spheroidal galaxies~\cite{McDaniel:2023bju}.
For both scenarios, the predicted fluxes with $g_\chi$ fixed by the GCE and antiproton excess remain below the corresponding observational sensitivities
and therefore do not constrain the favored parameter regions.

\begin{figure*}[!ht]
\centering
\includegraphics[width=0.49\textwidth]{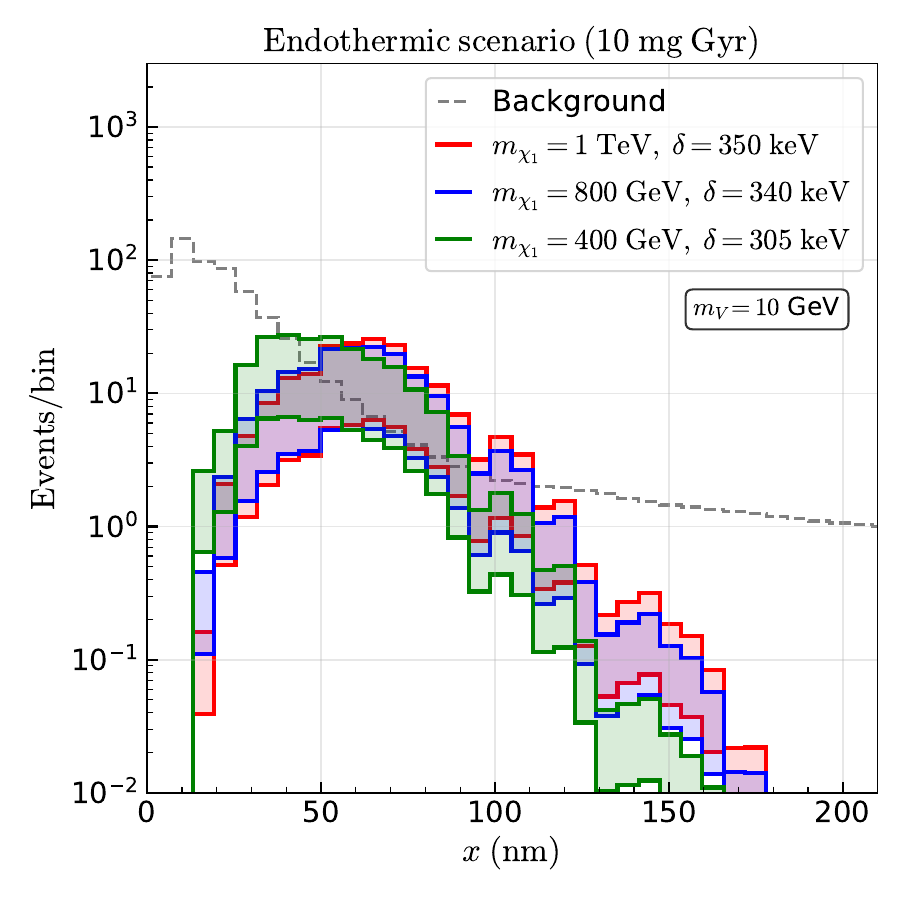}
\includegraphics[width=0.49\textwidth]{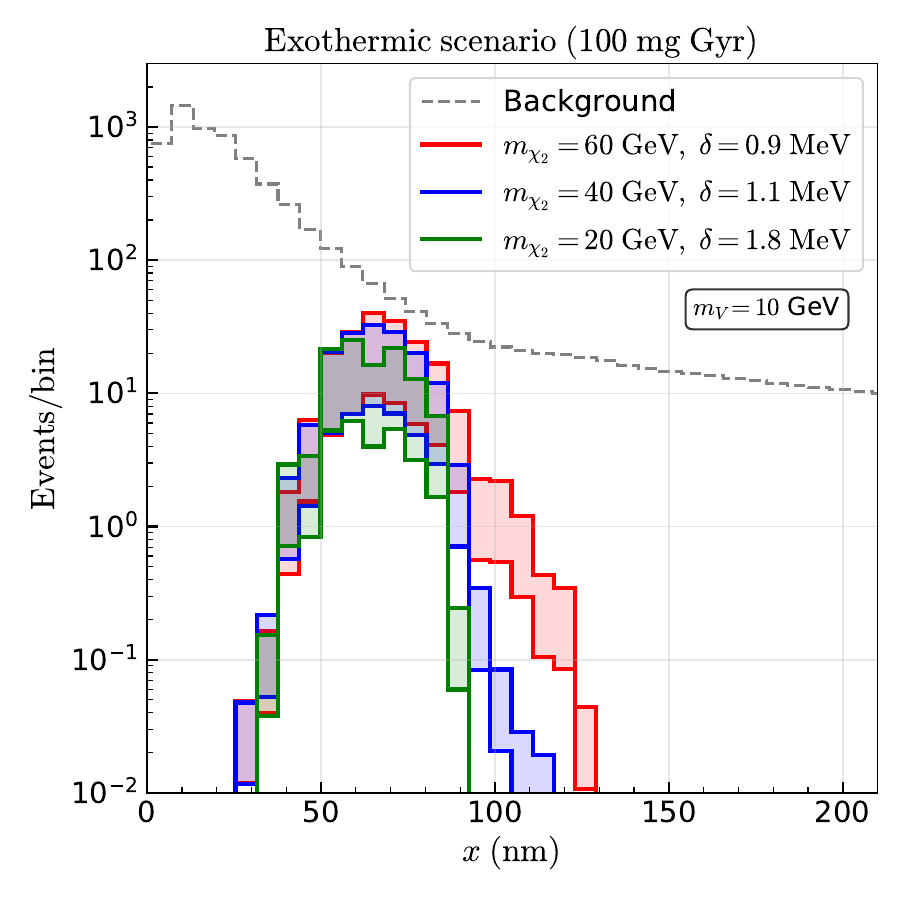}\\
\caption{Predicted track length distributions for the LZ anomaly against the background (gray) for $m_V=10\gev$.
Six benchmarks are same as Fig.~\ref{fig:signal},
with the width of each spectrum defined by the $g_\chi g_q$ signal region of the LZ anomaly reported in Ref.~\cite{LZ:2026axp}.
Assuming an optimistic $^{238}\mathrm{U}$ concentration of $10^{-12}\,\mathrm{g/g}$,
the signal can be confirmed with an exposure of $10~\mathrm{mg\,Gyrs}$ for the endothermic scenario (left) and
$100~\mathrm{mg\,Gyrs}$ for the exothermic scenario (right).}
\label{fig:Paleo}
\end{figure*}

\lettersection{Paleo-detector probe prospect}
The LZ anomaly fixes the inelastic mass splitting at $\delta\sim\mathcal{O}(300\,\mathrm{keV})$ for the endothermic scenario and $\delta\sim\mathcal{O}(\mathrm{MeV})$ for the exothermic scenario, as shown in Fig.~\ref{fig:signal}.
For light targets, the required threshold velocity $v_{\min}$ can exceed the halo escape speed,
rendering the scattering kinematically forbidden, particularly in the endothermic case.
A heavy target such as lead ($A=207$) substantially lowers $v_{\min}$ and thus enhances the accessible phase space.
Ancient lead-bearing minerals, such as laurionite, can permanently record the resulting damage tracks \cite{Graham:2026ivn},
providing a complementary probe of the LZ excess.

Following Ref.~\cite{Chu:2026skp}, we show six benchmark track-length spectra relative to the gray dashed background in Fig.~\ref{fig:Paleo}.
The endothermic and exothermic scenarios correspond to $10$ and $100$~mg~Gyr exposures, respectively, and follow the color scheme of Fig.~\ref{fig:signal}.
The characteristic DM-induced track length is $\sim 70$~nm, while the widths of the benchmark spectra denote the $90\%$ signal regions of $g_\chi g_q$ from Ref.~\cite{LZ:2026axp}.
For a $^{238}$U concentration of $10^{-12}$~g/g, the endothermic scenario is detectable with an exposure of only $10$~mg~Gyr,
whereas the exothermic scenario requires $\sim 100$~mg~Gyr.

The different exposure requirements arise from the distinct $v_{\min}$ regimes probed by the two scenarios.
For the $248\,\mathrm{keV}$ LZ ROI, endothermic scattering requires $v_{\min}\simeq750~\mathrm{km\,s^{-1}}$ for Xe,
deep in the high-velocity tail,
while Pb reduces this to $\simeq 650~\mathrm{km\,s^{-1}}$, substantially enlarging the accessible phase space.
For exothermic scattering, the corresponding values are only $\simeq 176$ and $114~\mathrm{km\,s^{-1}}$ for Xe and Pb, respectively,
both within the bulk of the halo distribution, thus the gain from the heavier target is much smaller.
Consequently, paleo-detectors can provide a particularly powerful test of the endothermic interpretation of the LZ excess.

If the LZ signal is confirmed to arise from exothermic scattering,
it would be consistent with the indirect-detection signals discussed above.
The endothermic scenario, in contrast, does not admit the same indirect-detection interpretation,
but can be independently tested with a paleo-detector exposure of only $\sim 10$~mg~Gyr.

\lettersection{Conclusion}
We have shown that leptophobic inelastic dark matter provides viable interpretations of the 248 keV LZ excess through both endothermic and exothermic channels. While the endothermic branch with $m_\chi\gtrsim400$ GeV cannot reproduce the Galactic Center gamma-ray excess,
the exothermic branch around $m_\chi\sim 55$ GeV can simultaneously account for the LZ excess,
the Galactic Center gamma-ray excess, and the AMS-02 antiproton excess.
The remaining slight relic-abundance tension can be alleviated by an additional neutrino-portal interaction without altering the quark-induced indirect signals.
Paleo-detectors provide an independent test of these interpretations,
with exposures of $\sim 10$ mg Gyr and $\sim 100$ mg Gyr potentially probing the endothermic and exothermic branches, respectively.

\lettersection{Note added}
While finalizing the revisions to our initial online preprint, Ref.~\cite{Gemmell:2026yaw} appeared, where the authors adopted an exothermic scenario to account for the LZ event and the GCE.
Relative to that study, the present independent work includes also the AMS-02 antiproton excess, and carries out an examination of the associated paleo-detector signatures.

\lettersection{Acknowledgments}
We were supported by
the National Natural Science Foundation of China (No. 12588101),
the National Key Research and Development Program of China (No. 2022YFF0503304),
the Project for Young Scientists in Basic Research of the Chinese Academy of Sciences (No. YSBR-092), the China Manned Space Program (No. CMS-CSST-2025-A03), and the Strategic Priority Research Program of the Chinese Academy of Sciences (No. XDB0550400).
We thank Daneng Yang and Xin-Chen Duan for their valuable help and insightful discussions during this work.

\bibliography{ref.bib}

\clearpage
\onecolumngrid
\begin{center}
    \LARGE
    \textit{Supplemental Material}
\end{center}
\setcounter{page}{1}
\renewcommand{\thepage}{S\arabic{page}}
\setcounter{secnumdepth}{1}
\setcounter{section}{0}
\renewcommand{\thesection}{S\arabic{section}}
\renewcommand{\theHsection}{S\arabic{section}}
\setcounter{equation}{0}
\renewcommand{\theequation}{S\arabic{equation}}
\renewcommand{\theHequation}{S\arabic{equation}}
\setcounter{figure}{0}
\renewcommand{\thefigure}{S\arabic{figure}}
\renewcommand{\theHfigure}{S\arabic{figure}}
\setcounter{table}{0}
\renewcommand{\thetable}{S\arabic{table}}
\renewcommand{\theHtable}{S\arabic{table}}
\vspace{1em}

\noindent
Table \ref{tab:benchmarks} summarizes the six benchmark points shown in Fig.~\ref{fig:signal} for $m_V = 10~\mathrm{GeV}$.

\begin{table*}[!htb]
\caption{Benchmarks of Fig.~\ref{fig:signal} for $m_V = 10~\mathrm{GeV}$, with $m_\chi = m_{\chi_1}$ for endothermic and $m_\chi = m_{\chi_2}$ for exothermic scattering.
The ranges of $g_q g_\chi$ are the $90\%$ C.L. signal regions of the LZ anomaly from Ref.~\cite{LZ:2026axp}, as in Fig.~\ref{fig:signal}, and $g_\chi$ follows from the relic density.
The heavy-state fraction $f_2$ is evaluated after the dark conversion and before any decay of $\chi_2$, using the Boltzmann solver of Ref.~\cite{Duan:2026zqj}.
}
\label{tab:benchmarks}
\begin{ruledtabular}
\begin{tabular}{llcc@{}c|cc}
\noalign{\vskip-\doublerulesep}
Scenario\rule{0pt}{2.6ex} & $m_\chi$ & $\delta$ & $g_q g_\chi$ (90\% C.L.) & & $g_\chi (\Omega h^2\approx 0.12)$ & $f_2 (\Omega h^2\approx 0.12)$ \\
\hline
Endo.\rule{0pt}{2.6ex} & $1~\mathrm{TeV}$ & $350~\mathrm{keV}$ & $(2.0\text{--}8.3) \times 10^{-5}$ & & $0.655$ & $0.019$ \\
 & $800~\mathrm{GeV}$ & $340~\mathrm{keV}$ & $(1.5\text{--}6.0) \times 10^{-5}$ & & $0.586$ & $0.021$ \\
 & $400~\mathrm{GeV}$ & $305~\mathrm{keV}$ & $(1.1\text{--}4.3) \times 10^{-5}$ & & $0.415$ & $0.122$ \\
\hline
Exo.\rule{0pt}{2.6ex} & $60~\mathrm{GeV}$ & $0.9~\mathrm{MeV}$ & $(1.2\text{--}4.8) \times 10^{-7}$ & & $0.159$ & $0.422$ \\
 & $40~\mathrm{GeV}$ & $1.1~\mathrm{MeV}$ & $(0.8\text{--}3.3) \times 10^{-7}$ & & $0.131$ & $0.474$ \\
 & $20~\mathrm{GeV}$ & $1.8~\mathrm{MeV}$ & $(0.53\text{--}2.2) \times 10^{-7}$ & & $0.096$ & $0.490$ \\[\doublerulesep]
\noalign{\vskip-\doublerulesep}
\end{tabular}
\end{ruledtabular}
\end{table*}

\section{Kinetic and mass mixing}
\label{sec:SM-mixing}

Besides Eq.~\eqref{eq:interaction} of the main text, the only renormalizable operators that mix $V_\mu$ with the neutral gauge bosons are the kinetic mixing with hypercharge \cite{Holdom:1985ag} and the mass mixing with the $Z$ boson \cite{Babu:1997st, Davoudiasl:2012ag}.
Below electroweak scale they read
\begin{equation}
    \mathcal{L}_{\mathrm{mix}} = -\frac{\varepsilon}{2} V_{\mu\nu} F^{\mu\nu} + \frac{\varepsilon s_W}{2 c_W} V_{\mu\nu} Z^{\mu\nu} - m_{VZ}^2 V_\mu Z^\mu,
    \label{eq:SM-mix}
\end{equation}
so that $V_\mu$ couples to the electromagnetic current ($e A_\mu J_\mathrm{EM}^\mu$ for the photon coupling) with strength $-\varepsilon e$ and, for $\abs{k^2} \ll M_Z^2$, to the weak neutral current with strength $-g_Z m_{VZ}^2 / M_Z^2$, where $g_Z = e / (s_W c_W)$.
The coupling of $V_\mu$ to the weak neutral current through $\varepsilon$ is further suppressed by $\abs{k^2} / M_Z^2$.

Quark loops generate $\varepsilon$, which runs at one loop as $\dv*{\varepsilon}{\ln \mu} = -(e N_c g_q / 6 \pi^2) \sum_q Q_q$ over the active flavors.
Below $m_c$ the charges of $u$, $d$, and $s$ sum to zero.
In the $\mathrm{SU}(3)$ flavor limit the vector current of Eq.~\eqref{eq:interaction} restricted to these quarks is a flavor singlet and their electromagnetic current is an octet, so the correlator of the two vanishes nonperturbatively; the breaking by $m_s$ leaves a finite residual of order $10^{-3} g_q$.
We therefore call $V_\mu$ leptophobic when the low-energy mixing $\varepsilon \equiv \varepsilon(\mu \lesssim m_c)$ is small.
This condition is stable under renormalization below $m_c$, because lepton loops renormalize $\varepsilon$ only multiplicatively.
Above $m_c$ the running regenerates
\begin{equation}
    \varepsilon(m_V) = -\frac{e N_c g_q}{6 \pi^2} \qty[\frac{2}{3} \ln\frac{m_b}{m_c} + \frac{1}{3} \ln\frac{m_V}{m_b}] \approx -0.017 \, g_q,
    \label{eq:SM-eps}
\end{equation}
and the full one-loop mixing at $q^2 = m_V^2$ gives $\abs{\varepsilon(m_V)} \approx 0.019\, g_q$.
Since $m_V$ lies $23~\mathrm{MeV}$ below the $\Upsilon(2S)$, the $b$-quark contribution is dominated by bottomonium, and the measured $\Upsilon(nS) \to e^+ e^-$ widths raise this to $\abs{\varepsilon(m_V)} \approx 0.07\, g_q$.
The resulting $\mathrm{BR}(V \to \ell^+ \ell^-) \approx 0.02 \abs{\varepsilon(m_V) / g_q}^2 \lesssim 10^{-4}$ is negligible, so $V$ decays into $q \bar{q}$ with $\mathrm{BR} \simeq 1$.
For $g_q \gtrsim 10^{-6}$ its lifetime is below $10^{-12}~\mathrm{s}$, which keeps the dark sector in equilibrium with the SM during freeze-out.

The mass mixing is not generated by Eq.~\eqref{eq:interaction}.
The quark current of Eq.~\eqref{eq:interaction} is conserved and its charge commutes with the flavor-diagonal $Z$ charges, so the Ward identity makes the $V$--$Z$ self-energy transverse.
A nonzero $m_{VZ}$ requires states that carry both charges and obtain mass from electroweak symmetry breaking, such as the fermions that cancel the anomalies of a gauged baryon number \cite{FileviezPerez:2010gw, Dobrescu:2014fca, Dobrescu:2021vak}.
With the SM fermions alone, the baryon-number current has the mixed anomalies $\mathcal{A}[\mathrm{SU}(2)_L^2 B] = -\mathcal{A}[\mathrm{U}(1)_Y^2 B] = 3/2$.
For $m_V > m_B - m_K \approx 4.8~\mathrm{GeV}$ the anomaly-enhanced decays $B \to K V$ and $K \to \pi V$ are closed, and Eq.~(7) of Ref.~\cite{Dror:2017ehi} gives $\mathrm{BR}(Z \to \gamma V) \approx 3 \times 10^{-8} \mathcal{A}^2 (3 g_q \mathrm{TeV} / m_V)^2 \approx 6 \times 10^{-3} g_q^2$.
This is at most $1.0 \times 10^{-10}$ for $g_q \lesssim 1.3 \times 10^{-4}$, far below the LEP bound for hadronic $V$ decays \cite{Dror:2017nsg}.

\section{Lifetime of the heavy state}
\label{sec:SM-lifetime}

For $\delta \ll m_{\chi_1}$ the decays of $\chi_2$ are Fermi-type transitions mediated by $V$ at $\abs{k^2} \lesssim \delta^2$, in which only the time component of the dark-matter transition current survives.
Through the mass mixing, summed over three neutrino flavors,
\begin{equation}
    \Gamma(\chi_2 \to \chi_1 \nu \bar{\nu}) = \frac{G_{\mathrm{eff}}^2 \delta^5}{40 \pi^3}, \qquad G_{\mathrm{eff}} = \frac{g_\chi g_Z m_{VZ}^2}{2 m_V^2 M_Z^2}.
    \label{eq:SM-nunu}
\end{equation}
The heavy state survives, $\Gamma t_0 < x$ with $t_0$ the age of the Universe, if
\begin{equation}
    m_{VZ} < \qty[\frac{160 \pi^3 M_Z^4 m_V^4\, x}{g_\chi^2 g_Z^2 \delta^5 t_0}]^{1/4},
    \label{eq:SM-mVZ}
\end{equation}
which for $x = 0.1$ gives $m_{VZ} < 2.8$, $2.4$, and $1.5~\mathrm{MeV}$ for the exothermic benchmarks with $m_{\chi_2} = 60$, $40$, and $20~\mathrm{GeV}$.
For endothermic scattering the residual $\chi_2$ must instead decay, because at equal coupling its exothermic down-scattering exceeds the endothermic LZ rate, computed with the standard halo model, the Helm form factor, and the LZ efficiency, by $2.0 \times 10^4$, $3.0 \times 10^4$, and $4.7 \times 10^4$ for $m_{\chi_1} = 0.4$, $0.8$, and $1~\mathrm{TeV}$.
Reducing the $f_2$ of Table \ref{tab:benchmarks} below the inverse of these ratios requires $\Gamma t_0 > 7.8$, $6.4$, and $6.8$, so $m_{VZ}$ must exceed the right-hand side of Eq.~\eqref{eq:SM-mVZ} with these values of $x$, namely $20$, $14$, and $13~\mathrm{MeV}$.

For $\delta > 2 m_e$ the kinetic mixing opens $\chi_2 \to \chi_1 e^+ e^-$ at tree level, with
\begin{equation}
    \Gamma(\chi_2 \to \chi_1 e^+ e^-) = \frac{(g_\chi \varepsilon e)^2 \delta^5}{60 \pi^3 m_V^4} \Phi(\delta),
    \label{eq:SM-ee}
\end{equation}
where $\Phi(\delta) = (30 / \delta^5) \int_{m_e}^{\delta - m_e} \dd{E}\, p p^\prime (E E^\prime - m_e^2)$ with $E^\prime = \delta - E$ accounts for the electron mass and equals $1.9 \times 10^{-3}$ and $0.27$ for $\delta = 1.1$ and $1.8~\mathrm{MeV}$.
Scattering on nuclei through Eq.~\eqref{eq:interaction} has the per-nucleon cross section $\sigma_n = 9 g_q^2 g_\chi^2 \mu_n^2 / (\pi m_V^4)$ for momentum transfers $\abs{\vb*{q}} \ll m_V$, where $\mu_n$ is the dark-matter--nucleon reduced mass; for $m_V = 10~\mathrm{GeV}$ this holds at the $\abs{\vb*{q}} \approx 250~\mathrm{MeV}$ of the LZ event.
Eliminating $g_\chi / m_V^2$ in favor of $\sigma_n$ gives
\begin{equation}
    \Gamma(\chi_2 \to \chi_1 e^+ e^-) = \qty(\frac{\varepsilon}{g_q})^2 \frac{e^2 \sigma_n \delta^5}{540 \pi^2 \mu_n^2} \Phi(\delta),
    \label{eq:SM-ee-sigma}
\end{equation}
so once the LZ rate fixes $\sigma_n$ the width depends on $\varepsilon / g_q$ but not on $m_V$.
The axial coupling of the $Z$ boson to electrons adds the mass-mixing contribution $\Phi_A(\delta) \Gamma(\chi_2 \to \chi_1 \nu \bar{\nu}) / 6$, where $\Phi_A$ is $\Phi$ with $+ m_e^2$ in the integrand.
Survival, $\Gamma t_0 < 0.1$, requires $\varepsilon / g_q < (0.95\text{--}3.9) \times 10^{-2}$ for $m_{\chi_2} = 40~\mathrm{GeV}$ and $(0.35\text{--}1.5) \times 10^{-3}$ for $20~\mathrm{GeV}$, where the ranges follow the intervals of $g_q g_\chi$ in Fig.~\ref{fig:signal} of the main text.

The Galactic 511 keV line gives a stronger condition.
Each decay injects a positron, which annihilates after slowing down.
We require the injection rate from $\chi_2$ within the inner $1.5$--$3~\mathrm{kpc}$ not to exceed the bulge annihilation rate $1.5 \times 10^{43}~\mathrm{s^{-1}}$ inferred from INTEGRAL/SPI \cite{Siegert:2015knp}, or twice that rate, for NFW, Einasto, and Burkert halos normalized to $0.4~\mathrm{GeV\,cm^{-3}}$ at the Sun.
This gives $\Gamma t_0 \lesssim 10^{-4}$ and hence $\varepsilon / g_q \lesssim (0.3\text{--}2) \times 10^{-3}$ and $m_{VZ} \lesssim 2~\mathrm{MeV}$ for $m_{\chi_2} = 40~\mathrm{GeV}$, and $\varepsilon / g_q \lesssim (0.9\text{--}6) \times 10^{-5}$ and $m_{VZ} \lesssim 0.5~\mathrm{MeV}$ for $20~\mathrm{GeV}$.
These decays realize mechanism (iii) of Ref.~\cite{Pospelov:2007xh}.
Near the bound they would contribute a subdominant component of the Galactic 511 keV emission, but a decaying source traces $\rho$ and is too extended to dominate it \cite{Vincent:2012an}, so we use the line only as a constraint.
Without a tree-level contribution, anomaly-cancelling fermions of mass $M$ would give $\varepsilon \approx (e N_c g_q / 6 \pi^2) \ln(M / 23~\mathrm{GeV}) \approx (0.02\text{--}0.06)\, g_q$ for $M = 0.1$--$1~\mathrm{TeV}$.
The benchmarks with $\delta > 2 m_e$ thus require the low-energy mixing to lie at least 10 and 400 times below this natural size, and for $m_{\chi_2} = 20~\mathrm{GeV}$ even below the hadronic residual, so the condition must be imposed on the physical low-energy mixing.
In particular, a leptophobic mediator, invoked in Ref.~\cite{Dent:2026bji} to allow $\delta > 2 m_e$, does not by itself close $\chi_2 \to \chi_1 e^+ e^-$.

\section{The gamma-ray and neutrino spectra of six benchmarks}
\label{app:gamma_nu}

\begin{figure*}[tb]
\includegraphics[width=0.49\textwidth]{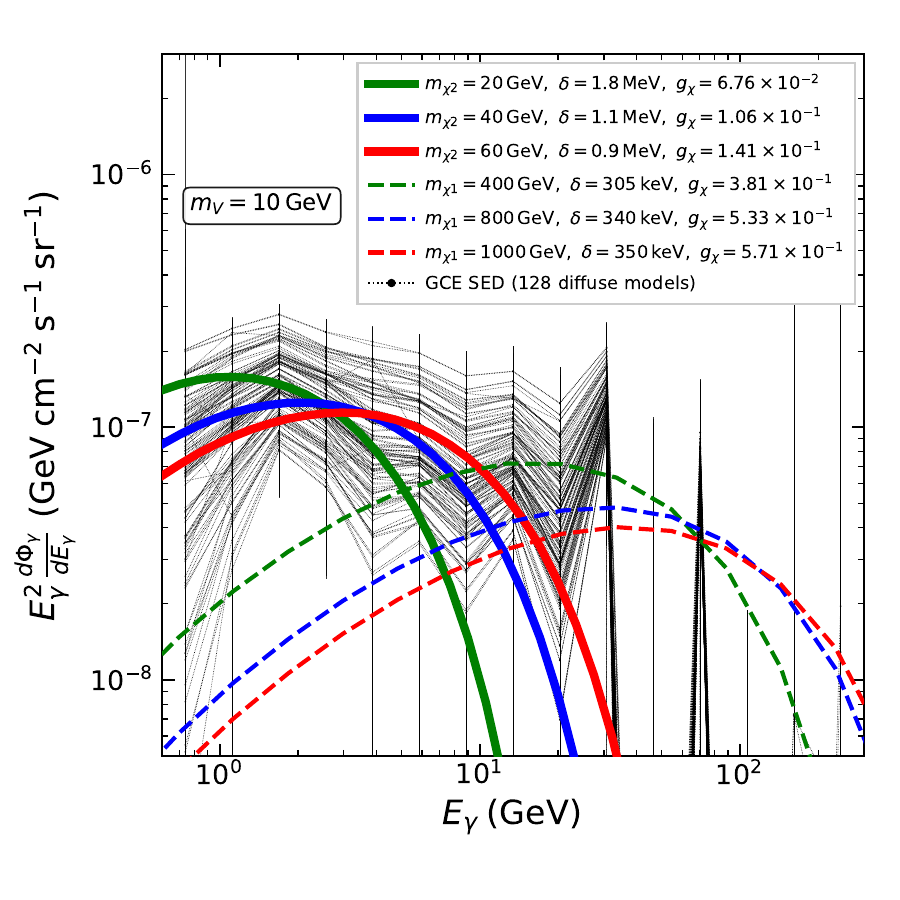}
\includegraphics[width=0.49\textwidth]{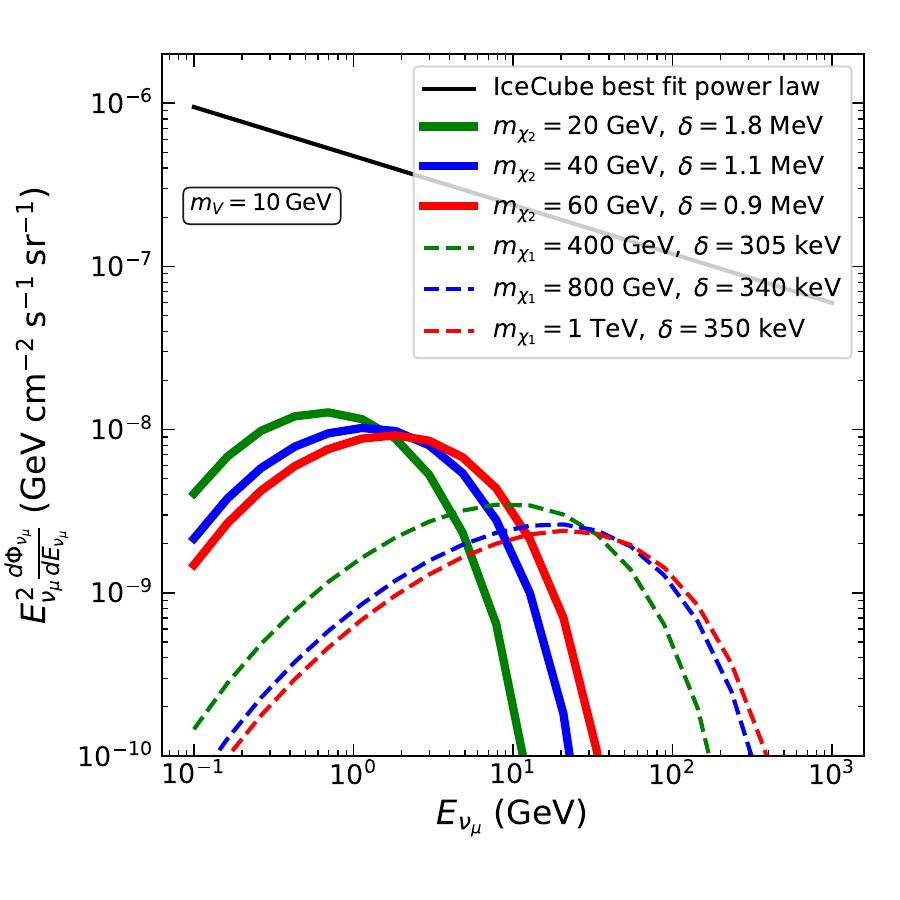}
\caption{DM annihilation gamma-ray and neutrino spectra for three exothermic benchmarks (solid lines) and endothermic benchmarks (dashed lines).
\texttt{Left:} GCE gamma-ray spectra for a gNFW profile.
The gray lines represent the uncertainties of 128 diffuse propagation models.
\texttt{Right:} Muon-neutrino spectra compared with the IceCube best-fit flux (black line).
The coupling $g_\chi$ is fixed by the GCE data and used to predict the neutrino flux.}
\label{app:ann}
\end{figure*}

Figure \ref{app:ann} presents our predicted gamma-ray spectra (left) compared to the Fermi-LAT Galactic center excess,
alongside the corresponding neutrino fluxes from the Galactic center (right).
The spectrum with the solid and dashed curves denoting the exothermic and endothermic scenarios, respectively.
Because the endothermic scenario requires a larger dark matter mass, its predicted spectral peak does not align with the observed GCE peak,
whereas the exothermic scenario with $m_\chi\approx 60\gev$ reproduces it well.
For both dark matter scenarios with $g_\chi$ fixed by the GCE data,
the corresponding neutrino fluxes (right panel) are an order of magnitude below the IceCube best-fit background model \cite{IceCube:2014stg},
making the dark matter contribution difficult to distinguish from the background with current data.
Future neutrino telescopes with improved sensitivity may still find it hard to distinguish a dark matter signal from the background.

\end{document}